\documentclass[sigconf]{acmart}

\AtBeginDocument{%
  \providecommand\BibTeX{{%
    \normalfont B\kern-0.5em{\scshape i\kern-0.25em b}\kern-0.8em\TeX}}}

\copyrightyear{2021} 
\acmYear{2021} 
\setcopyright{rightsretained} 
\acmConference[ESEM '21]{ACM / IEEE International Symposium on Empirical Software Engineering and Measurement (ESEM)}{October 11--15, 2021}{Bari, Italy}
\acmBooktitle{ACM / IEEE International Symposium on Empirical Software Engineering and Measurement (ESEM) (ESEM '21), October 11--15, 2021, Bari, Italy}\acmDOI{10.1145/3475716.3484194}
\acmISBN{978-1-4503-8665-4/21/10}

\include{commands}

\usepackage{threeparttable}
\usepackage{enumitem}
\usepackage{tabularx}
\usepackage[table]{colortbl}
\usepackage{graphicx}
\usepackage{siunitx}

\begin{document}
\setlength{\belowcaptionskip}{-10pt}

\title{Study of the Utility of Text Classification Based Software Architecture Recovery Method RELAX for Maintenance}

\author{Daniel Link}
\affiliation{
  \institution{University of Southern California}
  \streetaddress{8600 Datapoint Drive}
  \city{Los Angeles}
  \state{California}
  \country{USA}
  \postcode{78229}}
\email{dlink@usc.edu}

 \author{Kamonphop Srisopha}
 \affiliation{
   \institution{University of Southern California}
   \streetaddress{1 Th{\o}rv{\"a}ld Circle}
   \city{Los Angeles}
   \state{California}
   \country{USA}}
 \email{srisopha@usc.edu}

 \author{Barry Boehm}
 \affiliation{
   \institution{University of Southern California}
   \city{Los Angeles}
   \state{California}
   \country{USA}}
 \email{boehm@usc.edu}

\begin{abstract}

 \noindent \textbf{Background.} The software architecture recovery method RELAX produces a concern-based architectural view of a software system graphically and textually from that system's source code. The method has been implemented in software which can recover the architecture of systems whose source code is written in Java.
 
 \noindent \textbf{Aims.} Our aim was to find out whether the availability of architectural views produced by RELAX can help maintainers who are new to a project in becoming productive with development tasks sooner, and how they felt about working in such an environment.
 
 \noindent \textbf{Method.} We conducted a user study with nine participants. They were subjected to a controlled experiment in which maintenance success and speed with and without access to RELAX recovery results were compared to each other.
 
\noindent \textbf{Results.} We have observed that employing architecture views produced by RELAX helped participants reduce time to get started on maintenance tasks by a factor of 5.38 or more. While most participants were unable to finish their tasks within the allotted time when they did not have recovery results available, all of them finished them successfully when they did. Additionally, participants reported that these views were easy to understand, helped them to learn the system's structure and enabled them to compare different versions of the system. 
 
\noindent \textbf{Conclusions.} Through the speedup to the start of maintenance experienced by the participants as well as in their formed opinions, RELAX has shown itself to be a valuable help that could provide the basis of further tools that specifically support the development process with a focus on maintenance.
   
\end{abstract}

\begin{CCSXML}
<ccs2012>
   <concept>
       <concept_id>10011007.10010940.10010971.10010972</concept_id>
       <concept_desc>Software and its engineering~Software architectures</concept_desc>
       <concept_significance>500</concept_significance>
       </concept>
   <concept>
       <concept_id>10011007.10011074.10011081</concept_id>
       <concept_desc>Software and its engineering~Software development process management</concept_desc>
       <concept_significance>300</concept_significance>
       </concept>
   <concept>
       <concept_id>10011007.10011074.10011134.10011135</concept_id>
       <concept_desc>Software and its engineering~Programming teams</concept_desc>
       <concept_significance>500</concept_significance>
       </concept>
   <concept>
       <concept_id>10011007.10011074.10011111.10011696</concept_id>
       <concept_desc>Software and its engineering~Maintaining software</concept_desc>
       <concept_significance>500</concept_significance>
       </concept>
 </ccs2012>
\end{CCSXML}

\ccsdesc[500]{Software and its engineering~Software architectures}
\ccsdesc[300]{Software and its engineering~Software development process management}
\ccsdesc[500]{Software and its engineering~Programming teams}
\ccsdesc[500]{Software and its engineering~Maintaining software}

\keywords{Software Architecture Recovery, Text Classification, Software Maintenance}

\maketitle

\section{Introduction}

While several definitions of what a software architecture is exist \cite{solms2012software}, e.g., the set of design decisions about a software system \cite{Taylor2010}, they all refer to the structure of a software system and the reasoning process that led to that structure. A software architecture can be represented through many views that follow different paradigms, such as program comprehension and subsystem patterns \cite{tzerpos2000acdc}, optimized clustering \cite{Mancoridis1999}, dependencies, or concerns \cite{garcia2011enhancing,link2019recover}. (Note that the term \textit{concern} is used with the meaning \textit{something the system needs to have} and not \textit{something to worry about}.)

\begin{figure}
    \centering
    \includegraphics[width=0.9\columnwidth]{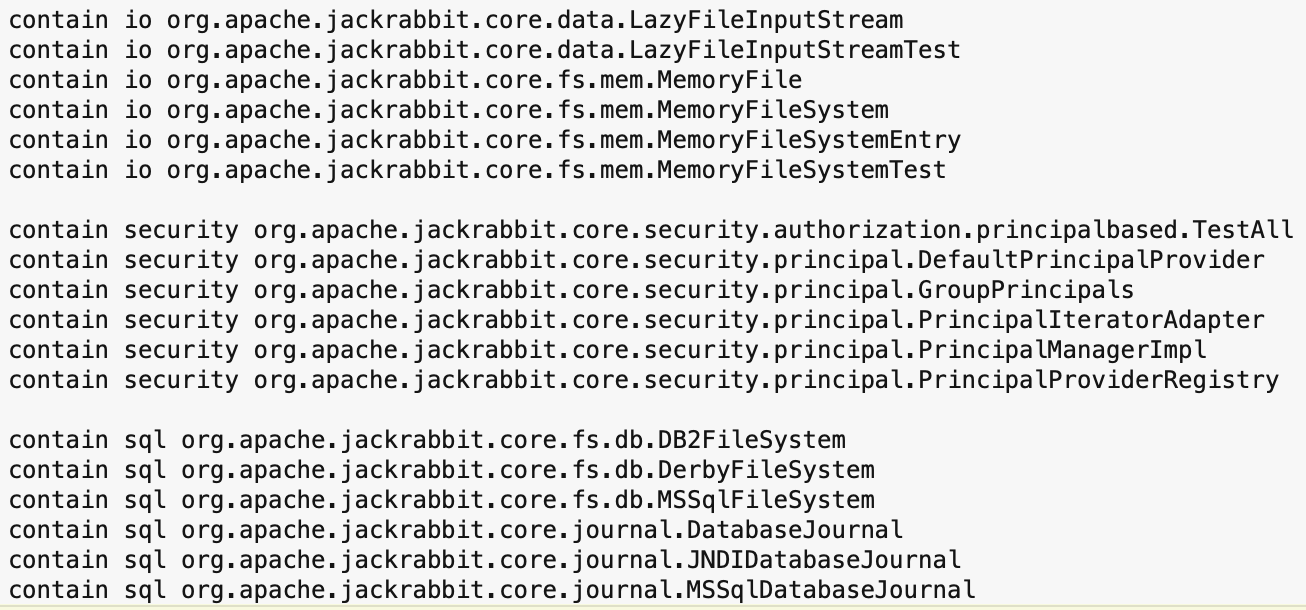}
    \caption{Partial Example of RELAX Textual Architecture Recovery Concern Clusters}
    \Description{Partial Example of RELAX Textual Architecture Recovery Concern Clusters}
    \label{fig:textresults}
\end{figure}

Having an actionable view of a software system is useful for its stakeholders for a variety of reasons, such as usability \cite{bass2003linking}, security \cite{bidan1997security}, and maintenance, as we are about to show. However, many reasons exist why an architectural view of a system that reflects its current state may not be available. These include that such a view may never have existed or that the system may have evolved away from it over time \cite{Taylor2010}. Therefore, an interest exists to produce such a view in an efficient and expedient manner. 
This is where software architecture recovery comes in. It produces an architectural view of a system from its implementation artifacts, such as its source code.
RELAX \cite{link2019recover} is a software architecture recovery method that follows a concern-oriented paradigm. It produces a software architecture by running text classification on the code entities of a system and building a view of the architecture from the results. The results are represented textually by a list of concern clusters (source code entities grouped by concerns) as shown in Figure \ref{fig:textresults} as well as graphically via a directory graph \cite{link2019recover}. Additional information for each source entity includes SLOC and dependencies.

From the outset, this method was designed with various groups of its end users in mind, such as system architects, developers, integrators and software engineering researchers, whether they would be interested in only one version of a system or plotting its architectural evolution over time. One important subgroup of developers is those who are new to a system, do not have access to current documentation (or any at all) and would like to get started with their work on the system as soon as possible, either because of approaching deadlines or in order to be seen as valuable contributors.

While several of RELAX's attributes, such as its determinism, efficiency, and scalability were able to be shown purely through tools that measured its runtime or sameness of results \cite{link2019recover}, its utility had not been studied yet. Specifically, we were interested in learning whether the task of getting started on modifying a system that was unknown to the study participants could be accelerated by having RELAX recovery results available.

The remainder of this paper is organized as follows: Section \ref{sec:approach} describes our approach. Section \ref{sec:results} details the results of our study. Section \ref{sec:discussion} examines the meaning of the results. Section \ref{sec:threats} addresses threats to its validity. Section \ref{sec:conclusion} sums up our conclusions, while Section \ref{sec:future} outlines our future work.

\section{Study Approach} \label{sec:approach}

The main objective of our study was to determine the utility of RELAX in a work environment.
A task that maintainers working with unfamiliar code encounter repeatedly and regularly is that of searching for relevant code to perform maintenance on \cite{Ko2006}. We concluded from this that if we could show that the availability of RELAX recovery results significantly reduces this time, it would hold utility for maintainers.
Accordingly, we simulated a situation in which individuals with at least some Java experience are assigned maintenance work on projects they are not familiar with and need to become productive as soon as possible.

\subsection{Study Questions} \label{subsec:studyquestions}

Our study sought to answer two research questions:

\begin{itemize}[label={}]
	\item \textbf{SQ1.} Does using RELAX architecture recovery results reduce the time to find the location in the code where maintenance needs to be performed?
	\item \textbf{SQ2.} What are the perceptions of new maintainers who work with RELAX architecture recovery results?
\end{itemize}

\subsection{Participants}

The software engineering group at our faculty holds several directed research sessions for graduate students of computer science which take place over the full duration of the regular semesters and Summer sessions. We sent out a call for participation in our forthcoming study to enrolled directed research students via email and held a presentation. In it, we outlined the project and made the main criterion outside of interest in the project that they should have Java programming skills commensurate with a bachelor’s degree in computer science. Prospective participants were offered extra credit. Nine participants joined our study.

\subsection{Duration}

Two considerations drove our selection of a duration for the study. On one hand, enough time needed to be available to instruct participants about software architecture, its recovery and RELAX. On the other hand, since the emphasis of the project was on studying performance and not educating the students, we could not expect to work with the participants for an entire semester. Our personal experience over several directed research projects had also shown us that the interest and with it the motivation wears out over time when tasks are repetitive and regarded as work. This would have risked influencing the outcome.

We felt that an overall duration of eight hours distributed over four weeks suitably reconciled the two opposing factors.

\subsection{Parameters}
During the four weeks our study ran, each participant contributed up to a maximum of two hours
per week, depending on whether and how fast they completed the tasks that were part of the study.
All nine participants from our convenience sample stayed for the entire duration.
The first week was spent on introducing 
them to or refreshing them on the basics of software architecture and instructing them on software architecture recovery with RELAX, followed by an initial survey on their experience levels going into the project.
The second week served to install, try out and validate the participants' hardware and software setups and perform dry runs with warm-up tasks that were not part of the study. The third week was spent exclusively on the first experimental task, while the fourth week was spent on the second task as well as the exit survey.

\subsection{Tools and Instrumentation} \label{subsec:toolsinstrumentation}

We gathered our data through surveys and observations.
To hold online surveys, the Qualtrics\footnote{qualtrics.com} platform was used. Live programming sessions were held over Zoom\footnote{zoom.us}. 

The participants had the option to choose between two 
maintenance environment setups, depending on their comfort levels:

\begin{enumerate}
\item A downloadable virtual machine that was pre-configured by us and contained a current desktop installation of Linux Ubuntu\footnote{ubuntu.com} with current versions of three major Java IDEs (Eclipse\footnote{eclipse.org}, NetBeans\footnote{netbeans.apache.org} and IntelliJ IDEA\footnote{jetbrains.com/idea/} as well as the current source code of Apache Jackrabbit\footnote{jackrabbit.apache.org} (version 2.20)
and Apache Ant\footnote{ant.apache.org} (version 1.9.15)
, which were the two systems that were chosen for the experiment.
\item Their own system with an IDE of their choice.
\end{enumerate}
The time to download source code and to set up the chosen IDEs was not counted in the second case, while in the first one this had already been done.
Time was recorded by us. The participants had 60 minutes to finish each task. A timer was started when the URL of the task description was sent to the participant via Zoom chat. 
If the participants finished a task successfully, time-taking was stopped and the actual time taken in minutes was recorded. If they reached the 60 minute mark without having finished the task, \textit{DNF} (standing for \textit{Did Not Finish}) was recorded and counted as 60 minutes.
In order to make timekeeping as accurate as possible, all tasks needed to be worked on live during a Zoom session in which the participants shared their screen and sound with the observer. This part of the setup aimed to prevent consultations of outside sources to the extent possible over a remote connection.
The observers were authorized to help the participants with their Zoom or Virtual Machine setups. However, such a need did not arise.

The open source Java projects Apache Jackrabbit and Apache Ant were chosen for the following principal reasons: 

\begin{itemize}
	\item No dependencies on older libraries or Java versions that may be hard to obtain
	\item Thousands of source files make it unlikely that participants can find maintenance locations by guessing or browsing
	\item Mature systems that are still updated and not likely to exhibit artifacts of antiquity or instability
\end{itemize}

\subsection{Tasks}

There were two warm-up tasks that were observed, but not counted, followed by two experimental tasks that were used for the study.

\subsubsection{Warm-Up Tasks} \label{subsubsec:warmups}

The warm-up tasks had the goal to establish the participants' programming skills and validate the infrastructure and setup of the experiment without using RELAX or its architecture recovery results.
We estimated these tasks to be feasible in less than ten minutes without knowing anything about the respective systems. The only items that were provided to the participants were the source code of the systems and instructions on how to run the command-line version of the systems in a shell.

The first, easier task was to change the textual output of Apache Jackrabbit on command-line startup from a shell. The second, more difficult task was to do the same with Apache Ant. While in the first case, this was achievable by making changes within the Java source code, it required modification of an external XML file in the second.

\subsubsection{Experimental Tasks} \label{subsubsec:experimentaltasks}

Since there were 9 participants, the split between the control group (participants not given the recovery results) and the experimental group (participants given the recovery results) could not be even, but was made 4 to 5 for the first task and 5 to 4 for the second (compare Table \ref{table:timing} below). Whichever participant was in the control group for the first task was in the experimental group for the second and vice-versa.

Both tasks were to be performed on Apache Jackrabbit 2.20 and had a common setting in which the participants were to assume that they know nothing about the architecture of Apache Jackrabbit
and are working alone in a closed and sealed bunker with no internet access. As surveyed at the outset of the study (Section \ref{subsec:experience}), none of the participants had any experience with the architecture or use of Jackrabbit and none had contributed any code to it. The participants had 60 minutes to finish each task. Timing was capped at 60 minutes as described in Section \ref{subsec:toolsinstrumentation}.\\

\noindent\textbf{Task 1:}

\begin{itemize}
    \item \textit{You are tasked with making fixes to a filesystem that interfaces with a database. In order to start your work, you think of adding a few print statements - but where?}
\end{itemize}

\noindent\textbf{Task 2:}

\begin{itemize}
    \item \textit{You are tasked to clean up system after an unreachable former contributor has made modifications to one source file so it does not handle security anymore, but SQL instead. The source file was not renamed. You want to make the system maintainable again and change the name of that file -- but where do you look?}
\end{itemize}

While the first experimental task was oriented toward an individual system version, the second one focused on the evolution of a system.

\subsection{Surveys}

The participants were given an initial survey in the first week, directly after the instruction phase (so that there would be no confusion over terms used) and an exit survey at the end of the study period.

The surveys consisted of a variety of question types, including 
text fields (e.g., for email addresses), multiple-choice (e.g., whether textual results, the diagram or both together gave participants the most complete information about the software system), yes-no (e.g., whether participants had contributed to Apache Jackrabbit), with rating questions shown as rating scales or Likert scales. 

To reduce fatigue and hold the attention of participants, the survey varied question types where possible. For example, some questions that could have been put on a Likert scale were put on ratings scales.
For the same reason, some ratings questions used sliders and some asked participants to enter their ratings as numbers in text fields. 

Questions are shown along with their results in Section \ref{subsec:exitsurveyresults}.

\subsubsection{Initial Survey (Participants' Experience)}

This asked participants about the levels of experience they had before the experiment started, regarding
\begin{itemize}
    \item Programming
    \item Text classification
    \item Software architecture and its recovery
    \item Apache Jackrabbit
\end{itemize}

Some questions, such as one question that asked participants whether they had contributed to Apache Jackrabbit or another that asked about preexisting experience with software architecture recovery opened further branches of questions on the relevant details. However, this did not apply to any of our participants.

\subsubsection{Exit Survey (Evaluations of RELAX)}

This asked the participants utility-related questions regarding RELAX recovery results and working with them on individual and multiple system versions.

\section{Results} \label{sec:results}

The full results of our study including the surveys are shared publicly \cite{https://doi.org/10.5281/zenodo.5323958}.

\subsection{Initial Survey on Preexisting Experience}\label{subsec:experience}

The top of Table \ref{table:survey} shows results of our survey on the experience the participants had before they began the study. Out of the fields surveyed, the only field they tended to have some experience in is that of text classification. They were new to Apache Jackrabbit when it came to its architecture and use and had next to no experience with its code and documentation. They had no experience with Software architecture recovery and none had contributed to Apache Jackrabbit.

Out of our nine study participants, eight had 1-5 years of programming experience, only one had 5-10 years of experience. Typically, they had some text classification experience, with the average being 2.56 on a scale of 1-5, with 5 being the highest.

Interpretation: This is in line with expectations for graduate students in computer science and was not likely to distort any results. It is particularly agreeable for our study that they did not have any significant experience with Apache Jackrabbit.

\begin{table} 
	\caption{Survey Results (Scales from 1-5)}
	\label{table:survey}
	\resizebox{\columnwidth}{!}{%
	\begin{threeparttable}
	\begin{tabular}{lcc}
		\toprule
		Question & Average & $\sigma$ \\
		\hline
		\cellcolor{lightgray!25}Familiarity with... &\cellcolor{lightgray!25} &\cellcolor{lightgray!25} \\
		Text Classification & 2.56 & 1.13\\
		Software Architecture Recovery & 1.00 & 0.00\\
		Apache Jackrabbit Source Code & 1.11 & 0.33\\
		Apache Jackrabbit Architecture & 1.00 & 0.00\\
		Apache Jackrabbit Documentation & 1.11 & 0.33\\
		Apache Jackrabbit Use & 1.00 & 0.00\\
		\hline
		\cellcolor{lightgray!25}RELAX view of Apache Jackrabbit: & \cellcolor{lightgray!25}& \cellcolor{lightgray!25}\\
		Easy to understand & 4.22 & 0.42\\
		I can learn the structure of the system from it & 4.78 & 0.42\\
		It can support new contributors & 4.67 & 0.47\\
		Real-time RELAX version valuable & 4.33 & 0.47\\
		\hline
		\cellcolor{lightgray!25}RELAX views of several Apache Jackrabbit versions: &\cellcolor{lightgray!25} &\cellcolor{lightgray!25} \\
		Easy to compare & 4.89 & 0.31\\
		Explain differences between versions & 4.89 & 0.31\\
		Describe system evolution & 4.67 & 0.47\\
		\hline
	
		\bottomrule
	\end{tabular}
	\begin{tablenotes}
	\item \small On the question whether the textual results, the diagram or both gave the most complete information on the architecture of a system, 3 participants chose the textual ones and 6 the combination of both.
	\end{tablenotes}
	\end{threeparttable}}
\end{table}

\subsection{Timings of Tasks}\label{subsec:timingtaskresults}

Table \ref{table:timing} shows how much time in minutes each participant took for each individual task. As mentioned, if they did not finish a task within 60 minutes, the time is shown as \textit{DNF} for ``Did Not Finish'' and counted as 60 minutes. The columns refer to the warm-up tasks described in Section \ref{subsubsec:warmups} and experimental tasks 1 and 2 
outlined in Section \ref{subsubsec:experimentaltasks}. 

The time RELAX took to recover an architectural view of Jackrabbit 2.20 on a 3.6 GHz Intel Core i9 using macOS was 235 seconds.
For both experimental tasks, the results are shown without and with counting the time the architectural view recovery took, rounded up to 4 minutes and added for each case in which a participant had the recovery results. Timings including the recovery runtime are shown in parentheses. The addition of this time is discussed in Section \ref{sec:discussion} below.

Before we look at the average speedups, two observations are notable: First, that of the nine participants, three were unable to complete the first warm-up task and six  the second. Second, while for each experimental task, only one participant was able to complete it without having recovery results available, all participants were able to complete their tasks when they had them.
Having recovery results has sped up the maintenance tasks by a factor of at least 5.38 for the first task or 7.44 for the second. These factors rise to 9.56 and 20.97 if we exclude the recovery time. It needs to be considered that in those cases where participants did not finish, the speedup factors are influenced by the timing cap of 60 minutes. Higher caps could have resulted in higher speedup factors.

\begin{table}
\caption{Task Timing Results (in Minutes)}
			\label{table:timing}
\resizebox{\columnwidth}{!}{
\begin{threeparttable} 
	\begin{tabular}{lrrrr}
		
		\toprule
		& \multicolumn{2}{c}{Warm-Up} & \multicolumn{2}{c}{Experimental}\\
		\cline{2-5}
		Participant & 1  & 2  & Task 1  & Task 2  \\
		\hline
		1 & DNF & DNF & DNF & \cellcolor{green!25}3.00 (7.00) \\
		\hline
	
		2 & DNF & 27.00 & DNF & \cellcolor{green!25}2.00 (6.00)  \\
		\hline
		
		3 & 11.50 & 8.00 & 6.00 & \cellcolor{green!25}2.00 (6.00)  \\
		\hline
	
		4 & 20.75 & 4.00 & DNF & \cellcolor{green!25}1.00 (5.00)  \\
		\hline
		
		5 & DNF & DNF & DNF & \cellcolor{green!25}3.00 (7.00) \\
		\hline
		6 & 12.50 & DNF & \cellcolor{green!25}7.08 (11.08) & DNF \\
		\hline
		7 & 14.50 & DNF & \cellcolor{green!25}2.50 \ \ (6.50) & 4.50  \\
		\hline
		8 & 16.50 & DNF & \cellcolor{green!25}5.00 \ \ (9.00) & DNF  \\
		\hline
		9 & 12.00 & DNF & \cellcolor{green!25}6.00 (10.00) & DNF  \\
		\hline
		&  &  &  &  \\
		\hline
		\midrule
		Averages & & & & \\
		\hline
		& 29.75 & 44.33 &  &  \\
		\hline
		Without recovery results & & & 49.20 & 46.13 \\
		\hline
		\rowcolor{green!25} With recovery results & & &  5.15 (9.15) & 2.20 (6.20) \\
		\hline
		Speedup factor & & & 9.56 (5.38) & 20.90 (7.44) \\
		\hline
		\bottomrule
	\end{tabular}
	\begin{tablenotes}
	\item \small Numbers in parentheses include RELAX runtime of 4 minutes.
	\item \small DNF (Did Not Finish) counts as 60 minutes.
	\end{tablenotes}
\end{threeparttable}
	}
\end{table}

\subsection{Exit Survey} \label{subsec:exitsurveyresults}

Table \ref{table:survey} shows the results of our exit survey in its lower two sections regarding working with either architectural views of one system version or views produced for several versions, respectively. The participants have formed overall favorable opinions from the view of one Jackrabbit version as well as the views of several versions. The highest rankings are held by the ease of understanding the view of a single version as well as comparing and explaining the differences between versions. One result that will be discussed below is that they found a hypothetical version of RELAX that would recover architectural views in the background valuable.

\section{Discussion} \label{sec:discussion}

The basis of our discussion can be summed up as follows:

\begin{itemize}
	\item The start of maintenance is sped up by at least a factor of 5.38
	\item Across the board, participants strongly considered architectural views produced by RELAX helpful in maintenance tasks
	\item This applies to both individual architectures and evolution
\end{itemize}

To which extent the recovery runtime for the current system version needs to be counted as part of the time the participant is working on a task depends on the specific setup regarding the number of maintainers and other factors. As discussed below in Section \ref{subsec:ideplugin}, we do not envision this to be a major timing factor.
However, we present our results with and without counting the runtime and conservatively report the longer time as our overall result.

\subsection{Answers to our Study Questions}
\subsubsection{\textbf{SQ1} (Does using RELAX architecture recovery results reduce the time to find the location in the code where maintenance needs to be performed?)}
The start of maintenance is sped up by factors of 5.38 and 7.44 for both tasks, respectively. Based on these results, SQ1 can therefore be answered affirmatively.

\subsubsection{\textbf{SQ2} (What are the perceptions of new maintainers who work with RELAX architecture recovery results?)}

As reported in Section \ref{subsec:exitsurveyresults}, the participants strongly considered the architectural views produced by RELAX to be helpful in their maintenance tasks. Based on this result, SQ2 can be answered affirmatively.

\subsection{Implications} 

\subsubsection{Enabling Maintenance}

Two standout observations come from Section \ref{subsec:timingtaskresults}. The first is that without having an architectural view of the system, for each of the two respective experimental tasks, only one participant was able to finish it. The second is that with an architectural view of the system, all 9 participants, including the 7 that had not successfully finished one or both warm-up tasks, were able to successfully finish their experimental tasks.
This means that even if one does not consider any speedups, a benefit of RELAX lies in making timely maintenance of a system possible.

\subsubsection{RELAX IDE Plugin} \label{subsec:ideplugin}

Based on the positive results of our study, in which we have demonstrated the utility of RELAX for maintenance tasks, we considered new tools for software maintainers, whether they are new to a system or know it well.
A possible tool would be a plugin for major IDEs.
One feature that sets RELAX apart from other recovery methods is that it assembles its architectural view in an additive manner from the text classification of individual source code entities of a system. This has the important consequences that individual source code files can be classified while they are being edited and that in order to see changes in the system, there is no need to recover the architecture of the whole system again because classifying the changed entities is sufficient.

We have considered how an IDE plugin that takes all this into account would be designed with regards to its internal and visual workflow. Some of the details of RELAX that the IDE would reference are beyond the scope of this paper and are described in \cite{link2019recover}. Four considerations would be that the plugin would (1) recover the system by request or automatically in the background after each successful compilation, (2) show the maintainer the current status of the architecture visually at a glance, (3) allow the kind of searches that helped the participants in our study, and (4) warn if the current architecture has issues, such as architectural smells \cite{Garcia2009}.

\begin{figure}[t]
	\centering
	\includegraphics[width=0.9\linewidth]{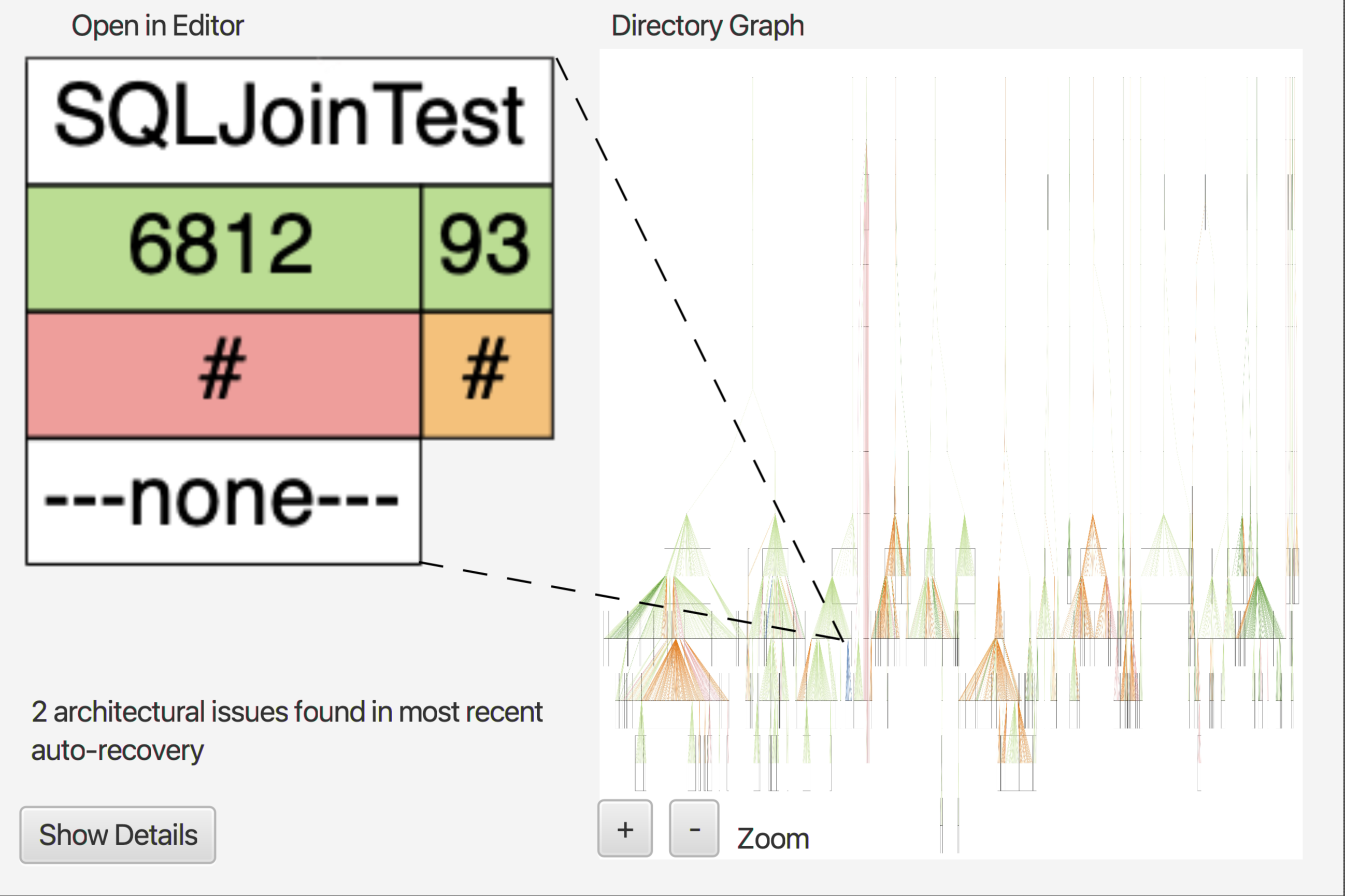}
	\caption[Status]{Plugin Status}
	\Description{Plugin Status}
	\label{fig:relaxidestatusmockup}
\end{figure}

\begin{figure}
	\centering
	\includegraphics[width=0.9\linewidth]{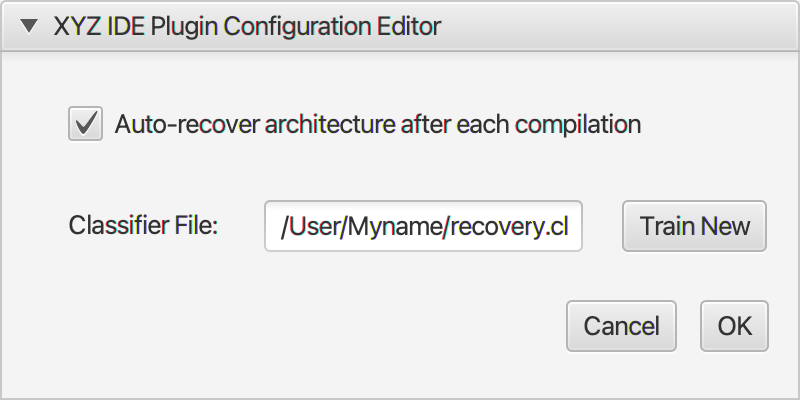}
	\caption[Configuratiion]{Plugin Configuration Dialog}
	\Description{Plugin Configuration Dialog}
	\label{fig:relaxideconfigurationmockup}
\end{figure}

\begin{figure}
	\centering
	\includegraphics[width=0.9\linewidth]{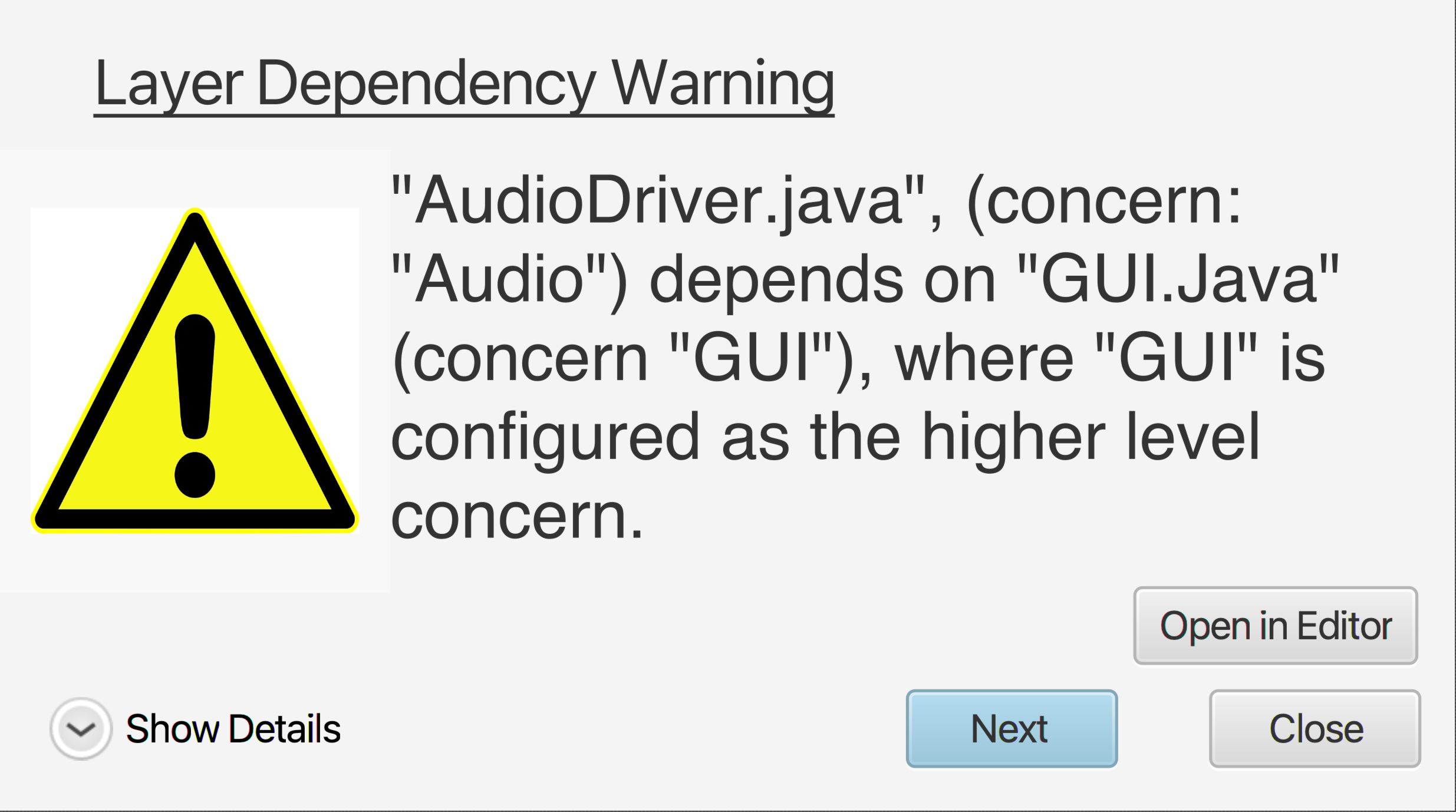}
	\caption[DepWarning]{Plugin Layer Dependency Warning}
	\Description{Plugin Layer Dependency Warning}
	\label{fig:relaxidelayerdependencywarningmockup}
\end{figure}

Figure \ref{fig:relaxidestatusmockup} shows the plugin status view that a maintainer could leave open in an IDE. The upper left shows information related to the current code entity being edited
, such as its name, file size, SLOC, color-coded outgoing and incoming dependencies as described in \cite{link2019recover}. The right half shows a graphical or textual view of the most recently recovered architectural view, 
allowing zooming in and panning. Finally, the lower left shows how many architectural issues have been found in the last recovery, 
along with a button that will open up a dialog showing issue details.
Figure \ref{fig:relaxideconfigurationmockup} shows the configuration dialog of the plugin. This would allow the maintainer to select or deselect auto-recovery after each compilation, select a trained classifier 
and to open a dialog that allows them to train a new classifier. The latter option could become helpful if the concerns of interest for the maintainer have changed.
Figure \ref{fig:relaxidelayerdependencywarningmockup} shows an example of an architectural warning that is enabled by the concern-orientation of RELAX. Maintainers could assign layers to some or all concerns in their system and the IDE plugin could analyze the architecture for layer breaches in which a lower-level entity calls a higher one. The maintainer could then check the details and take action by opening the file in the editor or view the next warning.

\section{Threats to Validity} \label{sec:threats}

\subsection{Task Selection}

The experimental tasks focused on finding code locations and not on other parts of the maintenance process, such as following dependencies, reading documentation, or more. Such activities would have been outside of the scope of our study, however.

\subsection{Subject System}

The subject system was selected after dismissing several other ones for not being a good fit due to lack of compilability, trivial size, not being open, not having any documentation (needed to come up with tasks) and other reasons.
\balance

\subsection{Number of Systems}

While a larger amount of systems might have provided more insights, this was not possible due to the limitations in the availability of participants.
In order to select a system that is as representative of a project that could conceivably take on new contributors, Apache Jackrabbit was chosen as project from the Apache Software Foundation\footnote{apache.org}, which hosts a large number of open source projects.

\subsection{Open Source vs Closed Source}

The programming style or quality of closed source systems may be significantly different from that of open source systems in ways that would lead to different results. 
However, studies \cite{spinellis2008tale,raghunathan2005open} have found no significant differences between the code quality of open- and closed-source systems.

\subsection{Java Programming Language}

A Java-based subject system was selected because the current version of RELAX only allows to recover architectural views from Java source code. The experimental tasks could be easier or more difficult if posed in different programming languages. Java, however is one of the most commonly used programming languages among developers worldwide \cite{liu_2021} and therefore 
a typical environment.

\subsection{Participants}

The participants were self-selected from computer science graduate students with at least some Java programming experience. While this group does not match up fully with all possible stakeholders in a system that can be served by RELAX recoveries, some of which can be non-technical, it does match up with the group of maintainers who would want to get started on their tasks.

\section{CONCLUSION} \label{sec:conclusion}

We conducted a user study with nine participants to assess the utility of the architectural recovery method RELAX. The participants were subjected to a controlled experiment in which speed and other metrics with and without access to RELAX recovery results were compared to each other on Apache Jackrabbit 2.20.

Through speeding up the start of maintenance by factors of at least 5.38 and other benefits experienced by the participants in our study, RELAX has shown itself to be a valuable help in the software maintenance process that can make the difference that allows new maintainers to finish a maintenance task on time.

\section{Future Work} \label{sec:future}

\subsection{Larger-Scale Study}
The encouraging emerging results from our study provide motivation for us to hold a larger-scale study with a broader scope.

Under consideration are several ways to achieve this.

\begin{itemize}
	\item \textit{Wider:} More systems and systems based on other programming languages than Java could be studied.
	\item \textit{Deeper:} More tasks could be given to the participants, either several on one system or on a number of systems.
	\item \textit{Larger:} A larger number of participants could be recruited and drawn from different pools, such as from the software industry and software engineers withe experience using other recovery methods.
\end{itemize}

\subsection{Software Support}

The envisioned IDE plugin that shows architectural changes as the system evolves  (see Section \ref{subsec:ideplugin}) could be developed together with former participants.

\bibliographystyle{ACM-Reference-Format}
\bibliography{relax-esem21}

\end{document}